\documentstyle[aps,twocolumn,epsfig,floats]{revtex}

\begin{document}  
\draft

 \wideabs{  
\author{M. G\"ockeler$^1$,            P.E.L.  Rakow$^1$,
            A. Sch\"afer$^1$,           W. S\"oldner$^1$, and
            T. Wettig$^{2,3}$}
\address{\medskip $^1$Institut f\"ur Theoretische Physik, Universit\"at
    Regensburg, D-93040 Regensburg, Germany\\ 
$^2$Center for Theoretical Physics, Yale University, 
New Haven, CT 06520-8120, USA\\
$^3$RIKEN BNL Research Center, Brookhaven National Laboratory, Upton, 
NY 11973-5000, USA} 
\date{March 23, 2001}
\title{Calorons and localization of quark eigenvectors in lattice QCD} 
\maketitle
   
 \begin{abstract}  
   We analyze the localization properties for eigenvectors of the
   Dirac operator in quenched lattice QCD in the vicinity of the
   deconfinement phase transition.  Studying the characteristic
   differences between the $Z_3$ sectors above the critical
   temperature $T_c$, we find indications for the presence of
   calorons.
 \end{abstract}
  }

One of the most discussed topics in hadron physics is the chiral phase
transition of QCD, and the microscopic processes connected with it.
Many current and planned experiments are at least partially motivated
by the hope that they will shed some light on the issues involved.

In this contribution we study the localization properties of the
lattice Dirac operator in the neighborhood of the chiral phase
transition. We are motivated by the fact that one of the most popular
pictures of the phase transition relates it to the properties of QCD
instantons and anti-instantons (see e.g.\ Ref.~\cite{Sha1}). The origin
of this connection is the following observation: For each isolated
instanton or anti-instanton there exists a localized zero mode of the
Dirac operator.  For a liquid of instantons and anti-instantons these
zero modes should be perturbed to form a band of small eigenvalues. At
higher temperatures it is thought that instantons and anti-instantons
may pair to form molecules, and that the associated modes will no
longer have particularly small eigenvalues, but instead become an
inconspicuous part of the bulk spectrum.

The Banks-Casher formula~\cite{BC},
\begin{equation}
\langle \bar q q \rangle = -\,{\pi\rho_{\rm Dirac}(0)\over V}\:,
 \label{BCrel} 
\end{equation}
relates the chiral condensate $\langle \bar q q \rangle$ to $\rho_{\rm
  Dirac}(0)$, the density of Dirac eigenvalues at zero, evaluated in
the (large) volume $V$.  Thus we expect that if the instantons and
anti-instantons form a liquid, chiral symmetry will be broken, but
that if they form instanton--anti-instanton molecules, chiral symmetry
will be restored.

If this interpretation is correct the chiral phase transition should 
reflect itself also in a characteristic change of the  localization
properties of the lowest Dirac eigenvectors. This is the main
motivation for our studies. 

In addition we are interested in comparing our lattice results with
predictions from random matrix theory (RMT). Such a comparison allows
us to distinguish generic from QCD-specific properties. Similar
studies were performed before in Ref.~\cite{Verb}. In this paper,
however, the authors analyzed a specific instanton-liquid
approximation to QCD while we study lattice QCD.

As usual, we simulate a finite-temperature system by working on a
lattice with $L_t < L_s$, where $L_t$ ($L_s$) is the temporal
(spatial) extent of the lattice. All boundary conditions are periodic
except for the temporal boundary conditions of the fermions which are
antiperiodic.  The temperature is given by $aT = 1/L_t$ with the
lattice spacing $a$.  We fix $L_t = 6$ and vary $a$ (and hence $T$) by
changing $\beta = 6/g^2$.

We work in the quenched approximation with staggered fermions, i.e.\
with the Dirac operator  
\begin{equation}
 D=  \sum_{\mu=1}^4  \frac{1}{2a} \alpha_{\mu}(x) \, 
 \left[\delta_{y,x+\hat \mu} U_\mu(x)
 - \delta_{y,x-\hat \mu}U_\mu^\dagger(y) \right] \:,
\end{equation} 
where $\alpha_{\mu}(x) = (-1)^{x_1 + \ldots + x_{\mu-1}}$ and the
$U_\mu$ are the link variables. The eigenvalues of $D$ come in pairs
of $\pm \lambda$, so we can restrict ourselves to positive eigenvalues
in the following. In the continuum limit this action corresponds to
four quark flavors. From now on we set $a$ to 1.

Quenched QCD has an additional $Z_3$ symmetry of the gauge sector,
which is spontaneously broken in the deconfined phase~\cite{Z3}. In
the confined phase the expectation value of the Polyakov loop $P$ is
zero, whereas in the deconfined phase $|P|$ acquires an expectation
value, and the phase of $P$ clusters around the values
$\theta_P$=arg($P$)=0,$\pm2\pi/3$. The fermion action does not share
the $Z_3$ symmetry, so fermionic quantities can depend on the $Z_3$
sector. It was found in Ref.~\cite{Chan} that the chiral condensate
computed from the configurations with $\theta_P =0$ vanishes above
$T_c$ as expected. For $\theta_P=\pm2\pi/3$ it remains finite in a
certain temperature range above $T_c$. This behavior can be understood
in Nambu--Jona-Lasinio models~\cite{Meis,Chan2} and in RMT~\cite{Ste}.
The point is that the boundary conditions of the Dirac operator are
not invariant under $Z_3$ transformations, and for
$\theta_P=\pm2\pi/3$ the new boundary conditions lead to a decrease of
the Dirac eigenvalues~\cite{Ste}. According to the Banks-Casher
relation, Eq.~(\ref{BCrel}), this implies that the condensate will
disappear only for larger temperatures, i.e.\ the transition
temperature is higher for these sectors. For phenomenological
comparisons one should use the $\theta_P$=0 sector as this one is
energetically favored for dynamical fermions. (The sectors
$\theta_P=\pm2\pi/3$ are physically equivalent to each other.)

Large temperatures correspond to a small extension of the lattice in
the temporal direction. The periodic boundary conditions for the gauge
fields thus imply that the field equations have solutions with
nontrivial topology which look rather like an instanton
chain~\cite{shepard}. Such configurations, called calorons, have been
the topic of intense investigations~\cite{Sha1}.

The index theorem tells us that the continuum Dirac operator has a
chiral zero mode in a caloron field, so we will search for calorons by
looking at the chiral and localization properties of the eigenvectors
of the low-lying eigenvalues of $D$. To calculate the eigenvalues and
eigenvectors we used the Arnoldi method as implemented in
Ref.~\cite{Arnoldi}. This method allows us to choose the number of the
lowest eigenstates and eigenvectors to be calculated.

\begin{figure}[t]
  \centerline{\epsfig{file=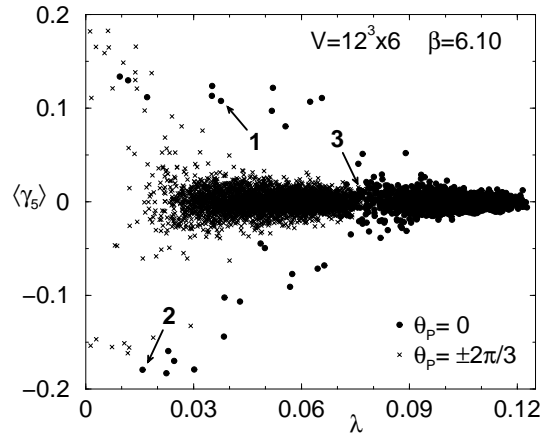,width=7cm}}
  \caption{ $\langle \gamma_5 \rangle$ vs.\ $\lambda$. 
    The measurements are on a $12^3\times 6$ lattice at $\beta=6.10$ for
    both sectors, $\theta_P=0$ and $\theta_P=\pm 2\pi/3$. 
    \label{g5plot}}
\end{figure}

\begin{figure}
  \centerline{\epsfig{file=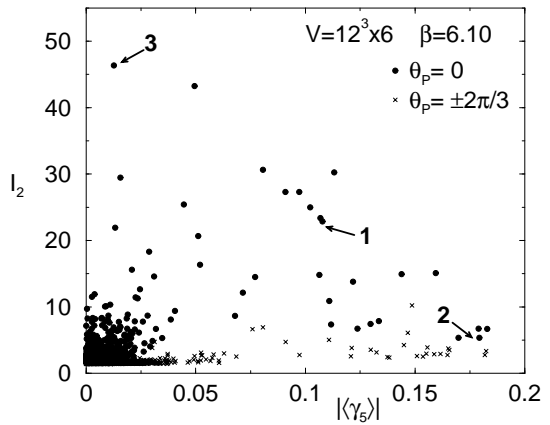,width=7cm}}
  \caption{ A scatter plot of the localization vs. chirality.
    \label{I2plot} } 
\end{figure}

In the continuum a zero mode has a definite chirality (= expectation
value $\langle \gamma_5 \rangle$) of $\pm1$, while modes with $\lambda
\neq 0$ have $\langle \gamma_5 \rangle = 0$. Staggered fermions
possess only a restricted chiral symmetry, so in this case we expect
$\langle \gamma_5 \rangle$ values between $-1$ and $1$. In
Fig.~\ref{g5plot} we show a scatter plot of $\langle \gamma_5 \rangle$
against Dirac eigenvalue, for a temperature slightly above $T_c$. We
can see that in all sectors the data points form clusters. There are
some eigenmodes with very small eigenvalue, and $\langle \gamma_5
\rangle \approx \pm 0.2$, while the bulk of the eigenvectors have a
larger $\lambda$, and $\langle \gamma_5 \rangle$ near zero. The modes
labeled 1 to 3 in Figs.~\ref{g5plot} and \ref{I2plot} will be
discussed below.

A large value of $\langle \gamma_5 \rangle$ suggests that the
corresponding eigenvectors have a topological origin, in which case
they should be associated with specific localized states. This
interpretation is supported by the fact that these modes are
approximately fourfold degenerate as they should be because of the
flavor symmetry of staggered fermions in the continuum limit. One of
our main results is that these states may be important for
understanding the differences between the $Z_3$ sectors.

As a measure of the localization of our quark eigenvectors
$\psi^\alpha_{\lambda}(x)$ ($\lambda$ is the Dirac eigenvalue,
$\alpha$ a color index) we use a gauge-invariant inverse participation
ratio (IPR), defined by
\begin{equation}
  I_2 \equiv V \frac{\sum_x p_{\lambda}(x)^2 }
  {\left[\sum_x p_{\lambda}(x)\right]^2}\:,
  \label{Idef} 
\end{equation} 
where $V$ is the number of lattice sites and $p_{\lambda}(x)$ is the
gauge-invariant probability density
\begin{equation} 
  p_{\lambda}(x) = \sum_{\alpha = 1}^{N_c} \left| \psi^\alpha_{\lambda}(x) 
  \right|^2 \,. 
  \label{rhodef} 
\end{equation}  
For a completely delocalized state (all $p_\lambda(x)$ the same) one
finds $I_2 =1$, whereas a state localized on a single lattice site (only one
non-zero $p_\lambda(x)$) would have $I_2 = V$. (The staggered
fermions' chiral symmetry  means that in fact an eigenstate can never
be completely localized, since all eigenstates must have half their
probability on even sites, and half on odd.) $I_2$ is also a measure
of the standard deviation of $p_\lambda(x)$,
\begin{equation}
  I_2 - 1 =  
  \frac{\sum_x \left[ p_{\lambda}(x) - \overline{p}_\lambda \right]^2}
  { V \overline{p}_\lambda^2 } \,. 
\end{equation} 
Chiral RMT predicts that the average value of $I_2$ is 
\begin{equation}
  \langle I_2\rangle=\frac{(N_c+1)V}{N_cV+2}\stackrel{V\to\infty}
  {\longrightarrow}1+\frac1{N_c} \quad{\rm for\;} N_c\ge3\:.
\end{equation}
To elucidate the relevance of the IPR let us note that in condensed
matter physics the size of the IPR decides e.g.\ whether a disordered
mesoscopic sample is a metal ($I_2$ close to the RMT prediction) or an
insulator ($I_2$ large).

We have investigated the behavior of $I_2$ on finite temperature
lattices, on both sides of the deconfinement phase transition, which
lies at $\beta \approx 5.89$ for $L_t = 6$ in the thermodynamic
limit~\cite{Biele}.

\begin{figure}[t]
  \centerline{\epsfig{file=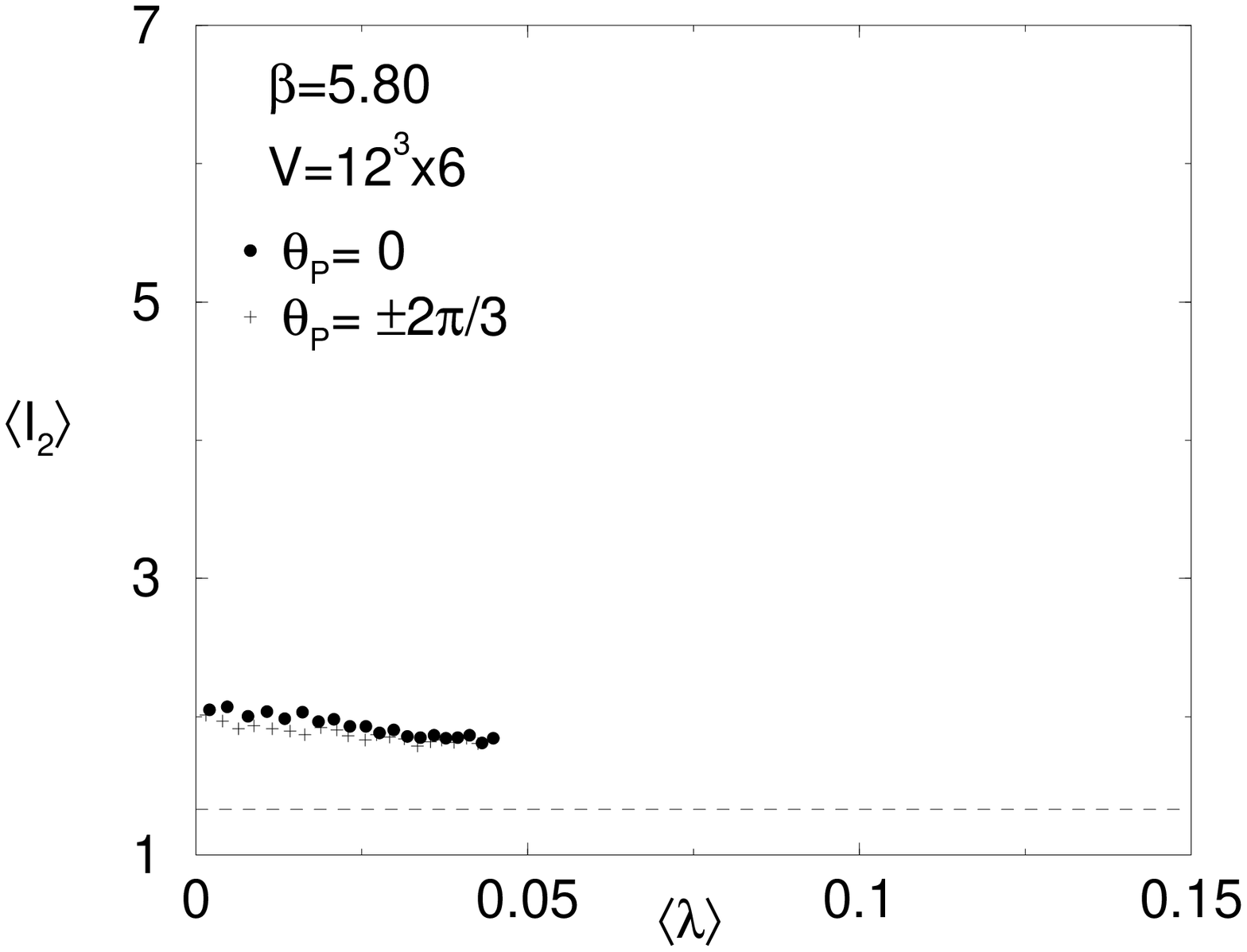,width=6.4cm}}
  \centerline{\epsfig{file=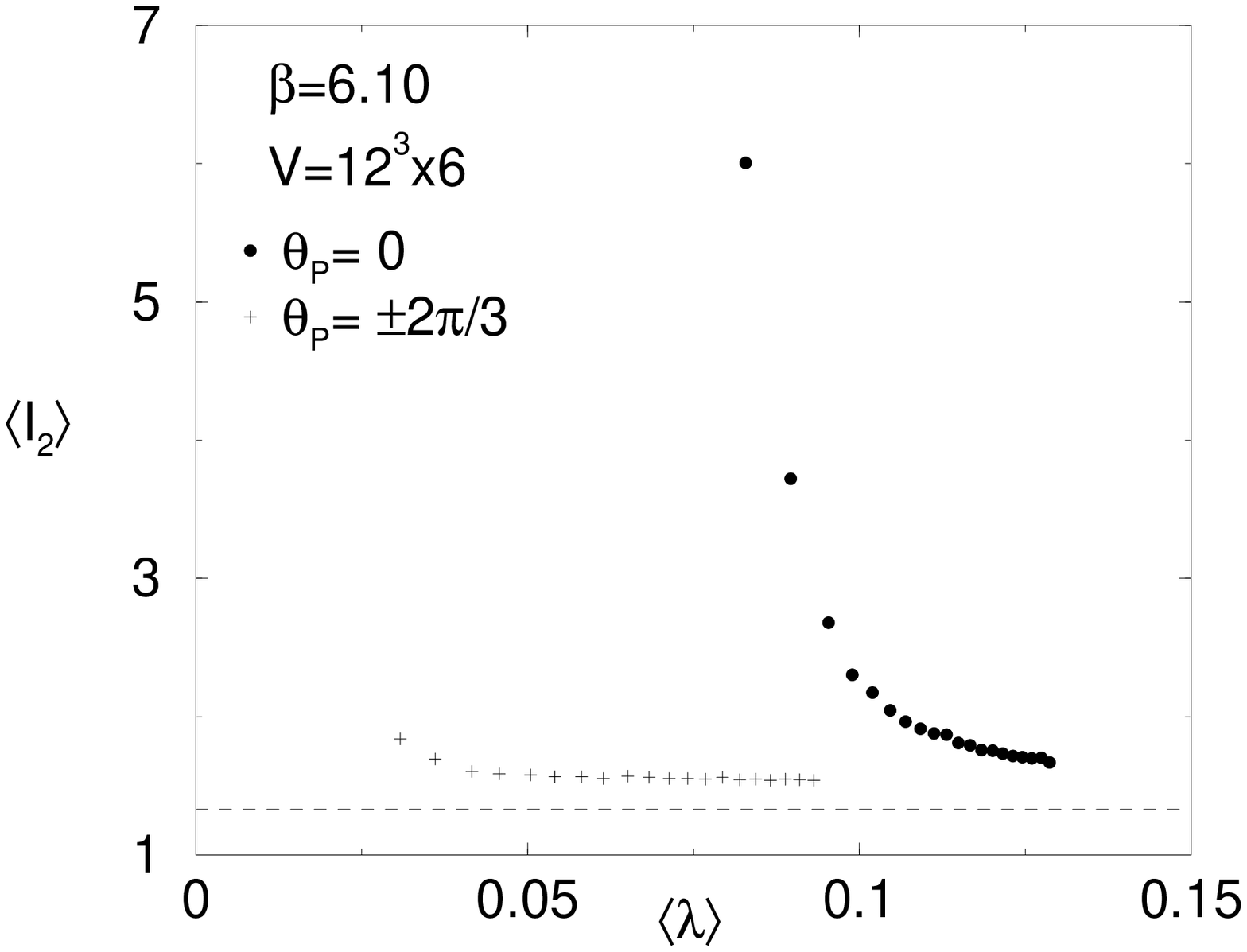,width=6.4cm}}
  \centerline{\epsfig{file=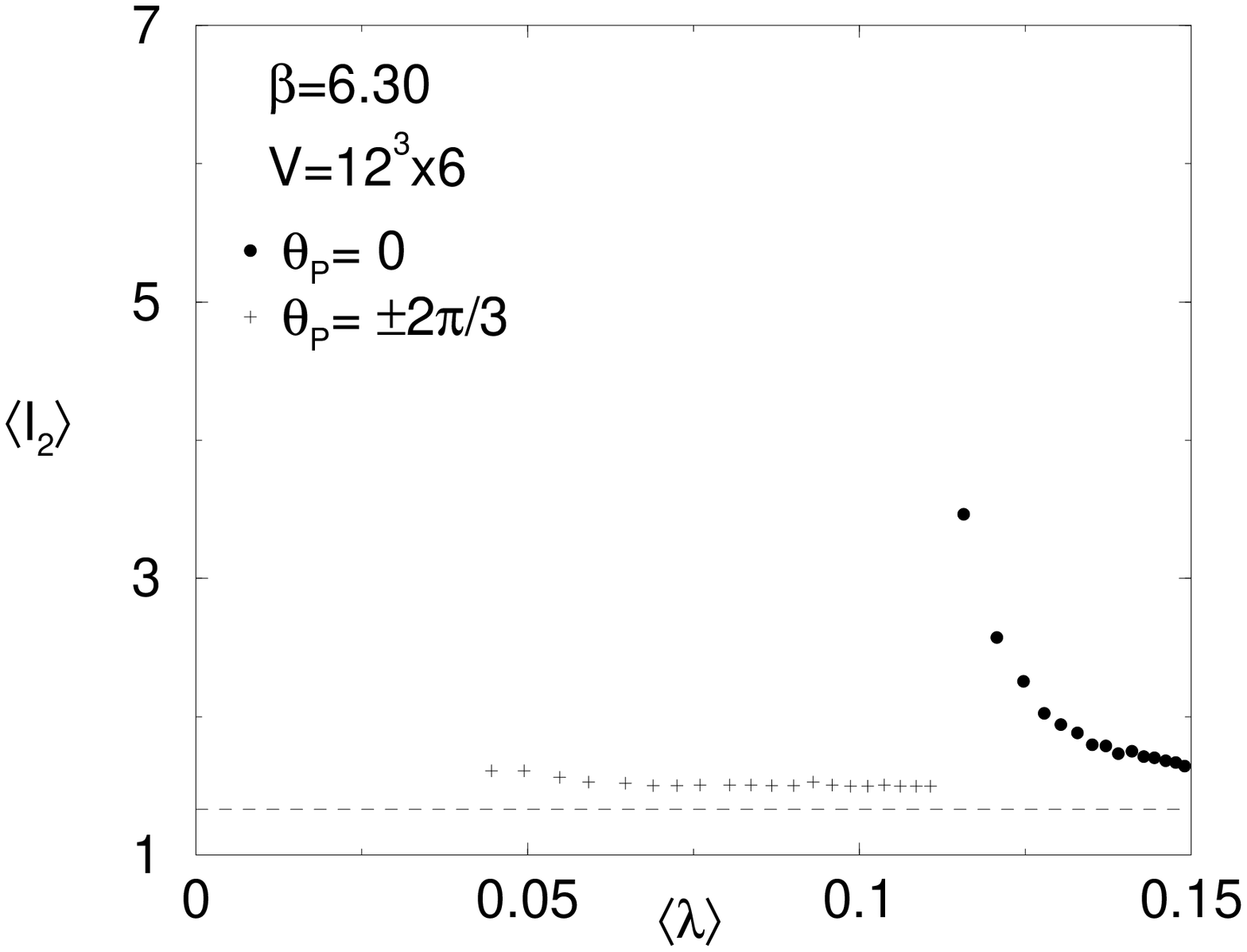,width=6.4cm}}
  \caption{ From top to bottom, 
    the average IPR for temperatures below, slightly above, and well
    above the chiral phase transition plotted separately for the $Z_3$
    sector with $\theta_P=0$ and the $Z_3$ sectors with $\theta_P=\pm
    2\pi/3$. The average was performed for the $1^{\rm st}$, $2^{\rm
      nd}$, \dots, $20^{\rm th}$ eigenvalue of each gauge field
    configuration. The dashed line is the RMT prediction, $4/3$.
    \label{fig3}}
\end{figure}

Slightly above $T_c$ the localization properties show characteristic
differences for the different $Z_3$ sectors. In the real sector the
effect of localization is strongly pronounced while it is nearly
absent in the $\theta_P=\pm 2\pi/3$ sectors where chiral symmetry is
still broken for these temperatures. This is illustrated in
Figs.~\ref{I2plot} and \ref{fig3}. A difference between localization
in the different sectors tells us immediately that the eigenstates
must be extended in the time direction, because an eigenstate
localized in time would be unaffected by a $Z_3$ transformation.

We studied these characteristic differences for high temperatures in
more detail and found features which suggest that they can be
attributed to the effects of calorons, as we shall now explain.

Harrington and Shepard found exact SU(2) instanton solutions at finite
temperature~\cite{shepard}, which they called calorons (for a review
see Ref.~\cite{review}). Their solution consists of a one-dimensional
chain of instantons, with a periodic repeat at a distance $1/T$. There
is a single physically relevant parameter, $R T$, which gives the
instanton radius $R$, relative to the scale set by the temperature.
When $R T \ll 1$ the solution looks like an isolated instanton, but
when $R T \sim 1$ finite temperature effects become significant. This
instanton chain satisfies the 't Hooft ansatz~\cite{multi}, so the
techniques described in Ref.~\cite{solve} can be used to find the
fermionic zero modes~\cite{Bilic}. For small $R T$ these look like the
zero modes of an isolated instanton, but as $R T$ increases, the modes
become extended in time, though they remain localized in space.  These
SU(2) solutions can easily be embedded in SU(3), leading to solutions
which correspond to the $\theta_P=0$ sector. To investigate the other
sectors, we need solutions where the time-like Polyakov loop has a
non-zero phase at space-like infinity. We can construct such solutions
either by adding a constant $A_4$ background field with a color that
commutes with the SU(2) subgroup containing the caloron, or by
choosing the fermion field boundary condition appropriately. In either
case the fermionic zero modes are readily constructed. One can also
find pure SU(2) solutions with a Polyakov loop
background~\cite{vBaal}. It would be interesting to study the
relationship between these solutions and the solutions found by
embedding the Harrington-Shepard solution in SU(3).

The localization of the embedded caloron's fermion zero mode depends
heavily on the $Z_3$ sector. Asymptotically the modes fall off like
\begin{equation}
  \left| \psi \right|^2 \propto \exp[ -2 (\pi - |\theta_P| ) \, r T ]/r^2
  \label{expon} 
\end{equation} 
where $r$ is the three-dimensional distance from the caloron axis. (At
$\theta_P =0$, this agrees with the behavior given in
Ref.~\cite{review}.)  This is unlike the case of a single instanton,
which has $|\psi|^2$ which drops off like a power of $r$. We see from
Eq.~(\ref{expon}) that the correlation length for the zero mode is
smallest in the real sector ($\theta_P =0$), and that the complex
$Z_3$ sectors ($\theta_P = \pm 2 \pi/3$) have modes in which the
radius is about three times greater, and so they occupy a much larger
volume. It is thus tempting to assume that the strong difference
between the localization properties in the $Z_3$ sectors is due to the
fact that our lattice configurations contain calorons. If so, the
localization in four dimensions should show the characteristic
string-like pattern discussed above.

\begin{figure}[t]
  \centerline{\epsfig{file=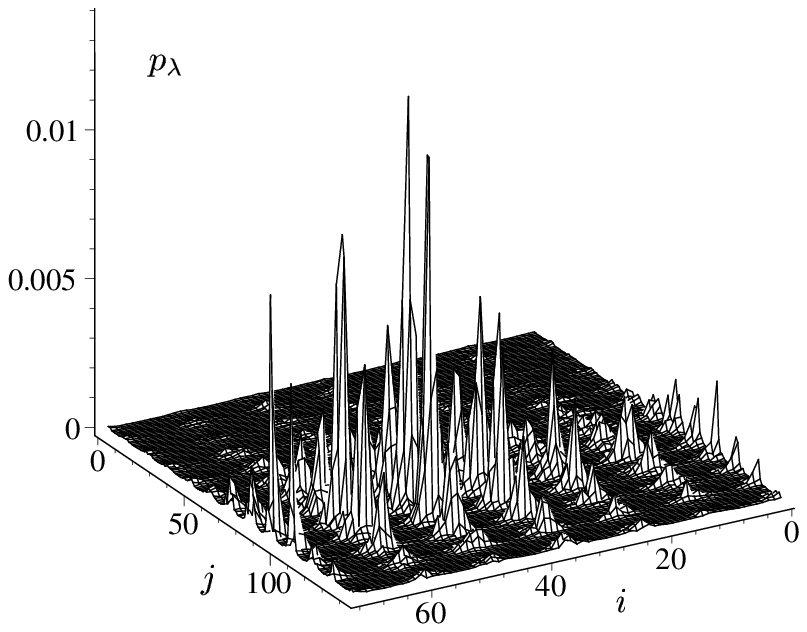,width=8cm}} 
  \centerline{\epsfig{file=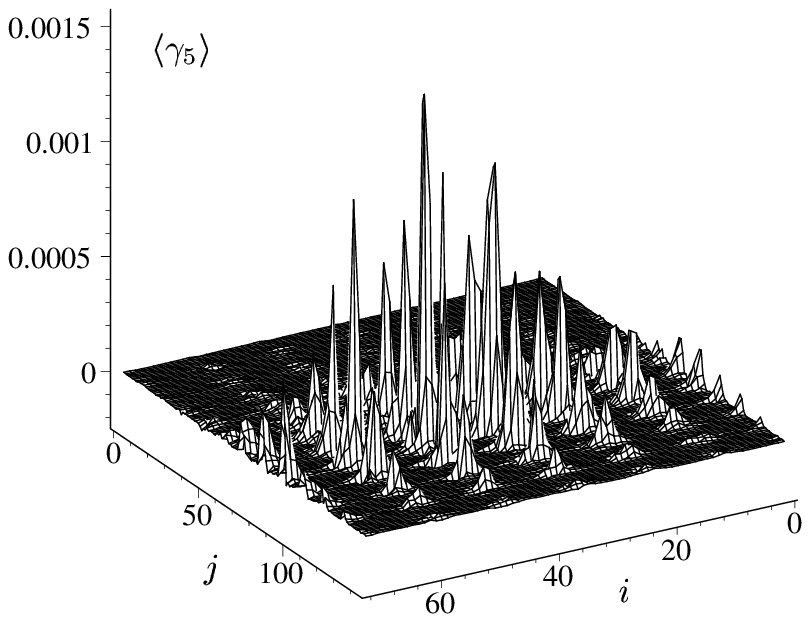,width=8cm}} 
  \vspace*{3mm}
  \caption{
    A `caloron' state (mode 1 of Figs.\ \ref{g5plot} and \ref{I2plot})
    on a $12^3\times6$ lattice for $\beta=6.1$ in the $\theta_P
    \approx 0$ sector. We have introduced the coordinates $i=x+12t$
    and $j=y+12z$ with the lattice coordinates $x,y,z=0,1,\ldots,11$
    and $t=0,\ldots,5$.  The gauge-invariant density is plotted above,
    the expectation value of $\gamma_5$ below.  Both show localization
    in space but not in time.
    \label{densit}}
\end{figure}

In Fig.~\ref{densit} the localization itself and the local density of
the expectation value of $\gamma_5$ are shown for a highly localized
state in the $\theta_P=0$ sector (mode 1 of Figs.~\ref{g5plot} and
\ref{I2plot}). To plot the complete four-dimensional lattice we have
introduced the two coordinates $i=x+12 t$ and $j=y+12 z$ with the
Euclidean lattice coordinates $x,y,z=0,1,2,\ldots,11$ and
$t=0,\ldots,5$. This coordinate system represents the lattice as a $6
\times 12$ array of $x$-$y$ slices, which generates the approximately
periodic structure visible in the plot. From the upper plot, we see
that the state is indeed spatially localized, but extended in time.
The lower plot demonstrates that the eigenstate in question is an
approximate chiral eigenstate. A continuum caloron would possess an
exact chiral eigenstate. Unfortunately, staggered fermions exhibit
only part of the continuum chiral symmetry, which is the reason why we
do not find perfect $\gamma_5$ eigenstates. It would therefore be very
interesting to repeat these studies with Ginsparg-Wilson fermions
which have much better chiral properties.

The corresponding pictures for mode 2 look similar to
Fig.~\ref{densit}. The width of the mode is, however, noticeably
larger than that of mode 1. Mode 3 serves as an example for a mode
with low $\langle \gamma_5 \rangle$, high $I_2$, and an eigenvalue
which lies in the bulk of the spectrum. Although $\langle \gamma_5
\rangle$ is small, the $\langle \gamma_5 \rangle$-density has a region
of large positive values right next to a region where the density is
large and negative. This suggests mode 3 might be an
instanton--anti-instanton molecule.

Modes related to instantons and calorons were recently investigated in
lattice QCD with dynamical fermions~\cite{Phil}. While the results
below $T_c$ are consistent with expectations based on an instanton
liquid picture, the interpretation of the observations above $T_c$
appears to be less straightforward.

Let us summarize: We have analyzed the localization properties of
quark eigenstates in quenched lattice QCD. By concentrating on the low
eigenvalues we could characterize semiclassical properties of the
gauge field configurations without any cooling. (For investigations
using cooling, see e.g.\ Ref.~\cite{cool}.) For temperatures above
$T_c$ we found isolated modes with definite handedness which show all
the properties of fermion states associated with calorons, in
particular localization in space but not in time.

We wish to thank A. V. Belitsky, Ph. de Forcrand, and T. Sch\"afer for
useful conversations.

\end{document}